\documentclass{ws-ijmpa}

\begin{document}


\markboth{V.~I.~Kuksa, R.~S.~Pasechnik, D.~E.~Vlasenko} {Mass shell
smearing effects in top pair production}

%
\catchline{}{}{}{}{}
%

\title{MASS SHELL SMEARING EFFECTS IN TOP PAIR PRODUCTION}

\author{V.~I.~KUKSA\footnote{kuksa@list.ru}}

\address{Institute of Physics, Southern Federal University, Rostov-on-Don 344090, Russia}

\author{R.~S.~PASECHNIK\footnote{roman.pasechnik@thep.lu.se}}

\address{Theoretical High Energy Physics, Department of Astronomy and
Theoretical Physics,\\ Lund University, SE 223-62 Lund, Sweden}

\author{D.~E.~VLASENKO\footnote{vlasenko91@list.ru}}

\address{Institute of Physics, Southern Federal University, Rostov-on-Don 344090, Russia}

\maketitle

\begin{history}
\received{Day Month Year} \revised{Day Month Year}
\end{history}

\begin{abstract}
The top quark pair production and decay are considered in the
framework of the smeared-mass unstable particles model. The results
for total and differential cross sections in vicinity of $t\bar{t}$
threshold are in good agreement with the previous ones in the
literature. The strategy of calculations of the higher order
corrections in the framework of the model is discussed. Suggested
approach significantly simplifies calculations compared to the
standard perturbative one and can serve as a convenient tool for
fast and precise preliminary analysis of processes involving
intermediate time-like top quark exchanges in the near-threshold
region.

\end{abstract}

\ccode{PACS number: 11.10.St}

\section{Introduction}

The top pair production and decay are the key processes for
precision tests of the Standard Model (SM) (see e.g.
Ref.~[\refcite{SM}] and references therein). They were intensively
studied in the framework of the Quantum Chromodynamics (QCD) and
Electro-Weak (EW) perturbation theory during last two decades, and
various methods and schemes were proposed. The major goal of these
investigations is to define the basic physical parameters of the top
quark, such as its mass, width and couplings with other SM
particles. In the past, the top quark physics was one of the primary
research objectives at Tevatron. Nowadays, the biggest attention is
paid to the process of the top quark production at the LHC (see e.g.
Refs.~[\refcite{1,2}]). However, the highest precision measurements
of the top quark properties can best be reached at the future Linear
Collider (LC) which supposedly operates in a clean experimental
environment. The top quark physics is one of the most interesting
and challenging targets for future $e^+e^-$ or $\mu^+\mu^-$ LC
experiments [\refcite{3,4}].

The top pair production is followed by a decay chain with
intermediate gauge boson states, i.e. the full process under
consideration is $e^+e^-\to t^*\bar{t^*}\to b\bar{b}W^+W^-\to
b\bar{b}4f$. The widths of both the top quark and the $W$-boson are
large, and one necessarily needs to take into account corresponding
Finite-Width Effects (FWE). In the framework of the standard
perturbative approach, these effects are typically described by
means of dressed propagators which are regularized by the total
decay width. In order to analyze the full process of the top pair
production relevant for phenomenological studies, we also have to
take into account the background contribution coming from many other
topologically different diagrams leading to the same six-fermion
final states, which is a rather non-trivial task.

The Born-level cross-sections of the processes $e^+e^-\to
b\bar{b}u\bar{d}\mu^-\bar{\nu}_{\mu}$ and $e^+e^-\to b\bar{b}4q$
were calculated in Refs.~[\refcite{5,6}] and [\refcite{7}],
respectively. Other exclusive reactions with
$b\bar{b}d\bar{u}\mu^+\nu_{\mu},\,b\bar{b}c\bar{s}d\bar{u}$ and
$b\bar{b}\mu^+\nu_{\mu}\tau^-\bar{\nu}_{\tau}$ final states were
considered in Ref.~[\refcite{4}]. In particular, it was shown that
the contribution of the top-pair signal $e^+e^-\to t^*\bar{t}^*\to
b\bar{b}4f$ is dominant, but the background (caused by one-resonant
or non-resonant diagrams) can be quite significant too. However, it
can be drastically decreased by applying certain kinematical cuts on
the appropriate invariant masses.

The QCD corrections for the reaction $e^+e^-\to t\bar{t}$ in the
continuum above the threshold were previously obtained in
Refs.~[\refcite{8,9}]. As well as the one-loop EW corrections were
calculated in many papers (for corresponding references, see e.g.
Introduction in Ref.~[\refcite{10}]). Concerning radiative
corrections (RC) to reaction $e^+e^-\to b\bar{b}4f$ with six-fermion
final states, the situation is more complicated and less clear
[\refcite{10}]. At the tree level, any of the reactions receives
contributions from several hundreds of diagrams. The calculations of
the full $O(\alpha)$ radiative corrections are very complicated, and
different approximation schemes are typically applied. The most
detailed analysis of the exclusive reactions $e^+e^-\to
b\bar{b}\mu^+\nu_{\mu}\mu^-\bar{\nu}_{\mu}$ and $e^+e^-\to
b\bar{b}d\bar{u}\mu^-\bar{\nu}_{\mu}$ was performed in
Ref.~[\refcite{10}]. In this paper, the cross-sections were
calculated taking into account the leading radiative corrections,
such as the initial state radiation (ISR) and factorizable EW
corrections to the on-shell top-pair production, to the decay of the
top quark into $bW$ and to the subsequent decays of the $W$-bosons.
Usually, such calculations are carried out automatically by Monte
Carlo techniques (see Ref.~[\refcite{10}] and references therein).

In this work, we consider reactions like $e^+e^-\to t^*\bar{t}^*\to
b\bar{b}4f$ with any four-fermion final states $4f$. The analysis is
performed in the framework of the smeared mass unstable particles
model (below, SMUP model) [\refcite{11,12}]. Due to exact
factorization at intermediate $t,\bar{t}$ and $W^+,W^-$ states, the
cross-section can be represented in a simple analytical form which
is convenient for analytical and numerical analysis. So far, we have
applied the SMUP approach only for unstable gauge boson production
and decay (see e.g. Refs.~[\refcite{13,14,14a}]). As a continuation
of our earlier studies, in this work we test the SMUP approach for
the case of unstable fermions, i.e., specifically, top quarks. In
our calculations, we take into account NLO radiative EW and QCD
factorizable corrections which dominate close to $t\bar{t}$
threshold. Also, we illustrate the influence of the mass smearing
effects and various radiative corrections (RC's) on the differential
cross-sections. The results are compared with ones calculated by
using the standard perturbative methods [\refcite{10}], where
cross-sections were represented for case of full $2\to 6$ process
and, separately, for the top signal contribution alone. It was shown
that in the Born approximation the results coincide with a rather
high precision, and deviations of the higher-order corrected results
from the standard ones are at the percentage level. So, the
suggested approach can be applied in a fast preliminary analysis of
various complicated processes involving intermediate top quark
exchanges in the Standard Model and beyond.

Note, here we do not consider the near-threshold effects caused by
the generation of the coupled $t\bar{t}$ state, which were
considered in detail in many previous studies (see, for instance,
Ref.~[\refcite{15}] and references therein). We postpone this issue
for a forthcoming study.

\section{The model cross-section of the top-pair production and decay at the tree level}

\begin{figure}
\centerline{\includegraphics[width=.6\textwidth]{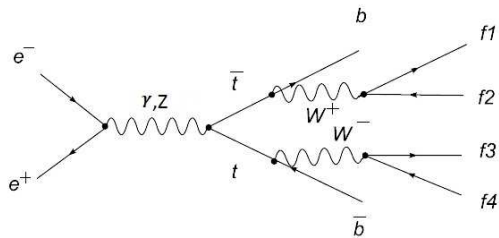}}
\caption{Feynman diagram of the top quark signal process $e^+e^-\to
t^*\bar{t^*}\to b\bar{b}W^+W^-\to b\bar{b}4f$.} \label{fig1}
\end{figure}

The process of top-pair production with subsequent decay $e^+e^-\to
t^*\bar{t^*}\to b\bar{b}W^+W^-\to b\bar{b}4f$ is schematically
represented in Fig.~\ref{fig1}. The full process contains two steps
with unstable intermediate time-like states, namely, $t,\bar{t}$ and
$W^+,W^-$ states. In this case, as was shown in Ref.~[\refcite{12}],
the double factorization takes place and can be described in the
framework of the SMUP model [\refcite{11}]. Due to this
factorisation, the full process can be divided into three stages:
$e^+e^-\to t^*\bar{t}^*$, $t^*\bar{t}^*\to b\bar{b}W^+W^-$ and
$W^+W^-\to 4f$. Here, the top-quarks and $W$-bosons are treated as
unstable particles, and finite-width effects should be taken into
account.

The SMUP model cross-section of the first reaction $e^+e^-\to
t^*\bar{t}^*$ can be written as [\refcite{11}]
\begin{equation}\label{2.1}
\sigma(e^+e^-\to
t^*\bar{t}^*)=\int_{m^2_0}^s\int_{m^2_0}^{(\sqrt{s}-m_1)^2}
\sigma(e^+e^-\to t(m_1)\bar{t}(m_2))\rho_t(m_1)\rho_t(m_2)dm^2_1
dm^2_2,
\end{equation}
where $m_0\approx 2M_b$ ($M_b$ is the bottom quark mass) is the
threshold value of the top mass variable, $\sigma(e^+e^-\to
t(m_1)\bar{t}(m_2))$ is the cross-section of top pair production
with random masses $m_1$ and $m_2$ and $\rho_t(m)$ is the
probability density which describes the mass smearing of top quarks.
In our calculations we take it in the Lorentzian form as
[\refcite{11}]
\begin{equation}\label{2.2}
\rho_t(m)=\frac{1}{\pi}\frac{m\Gamma_t(m)}{(m^2-M^2_t)^2+m^2\Gamma^2_t(m)},
\end{equation}
where $\Gamma_t(m)$ is the total decay width of the top quark with
mass $m$. The decay mode $t\to bW$ has a very large branching ratio
$\mathrm{Br}(t\to bW) \approx 0.999$, so formula (\ref{2.1}) almost
exactly describes the cascade process $e^+e^-\to t^*\bar{t}^*\to
b\bar{b}W^+W^-$ in the stable $W$-boson approximation. In order to
take into account the instability of $W$-bosons we have to express
the top quark width $\Gamma_t(m)\approx \Gamma(t\to bW)$ in
Eq.~(\ref{2.2}) as a function of smeared $W$-boson mass $\Gamma(t\to
bW(M_W))$ with averaging over $M_W$. Thus, the model cross-section
of the full inclusive process $e^+e^-\to t^*\bar{t}^*\to
b\bar{b}W^+W^- \to b\bar{b}\sum_f 4f$ depicted in Fig.~\ref{fig1}
has the following convolution form:
\begin{align}\label{2.3}
\sigma(e^+e^-\to b\bar{b}\sum_f
4f)=&\int_{m^2_0}^s\int_{m^2_0}^{(\sqrt{s}-m_1)^2}
\sigma(e^+e^-\to t(m_1)\bar{t}(m_2))\times \notag\\
&\int_{(m_0-M_b)^2}^{(m_1-M_b)^2}\rho_t(m_1,m_{W^+})\rho_W(m_{W^+})dm^2_{W^+}\times\\
&\int_{(m_0-M_b)^2}^{(m_2-M_b)^2}\rho_t(m_2,m_{W^-})\rho_W(m_{W^-})dm^2_{W^-}dm^2_1
dm^2_2,\notag
\end{align}
where $\rho_W(m)$ is defined by Eq.~(\ref{2.2}). In order to
describe an exclusive reaction $e^+e^-\to t^*\bar{t}^*\to
b\bar{b}f_1f_2f_3f_4$ we have to replace the total decay widths of
$W$-bosons, which enter the numerator in Eq.~(\ref{2.2}), by
corresponding exclusive ones (see Section 4). The same result can be
obtained exactly if one calculates the cross-section of this process
explicitly in the framework of the SMUP model by using dressed
propagators of unstable particles (UP's). In Ref.~[\refcite{12}] it
was shown that exact factorization of a decay chain process with
UP's in an intermediate state takes place when we exploit the model
effective propagators for fermion and vector UP's in the following
form
\begin{equation}\label{2.4}
\hat{D}(q)=i\frac{\hat{q}+q}{P_F(q)},\qquad
D_{\mu\nu}(q)=-i\frac{g_{\mu\nu}-q_{\mu}q_{\nu}/q^2}{P_V(q)},
\end{equation}
where $P_F(q)$ and $P_V(q)$ are the denominators of the fermion and
vector boson dressed propagators, which contain corresponding total
decay widths. The structure of numerators in Eq.~(\ref{2.4})
provides exact factorization and leads to a convolution-like
expression (\ref{2.3}) for the cross-section. So, there is a
self-consistency between the model and UP effective theory
description of the processes with UP in intermediate states. Thus,
the process with a six-particles final state shown in
Fig.~\ref{fig1} is described by a simple analytical expression
(\ref{2.3}) with four integrations over smeared unstable top and
$W$-boson masses. Note, that the standard perturbative treatment of
the six-particle final states in general case leads to $N=3\cdot
6-4=14$ independent parameters, from which 13 parameters have to be
integrated over [\refcite{16}]. Such a complicated problem can be
solved by using involved Monte Carlo numerical simulations only.

\begin{figure}
\centerline{\includegraphics[width=.6\textwidth]{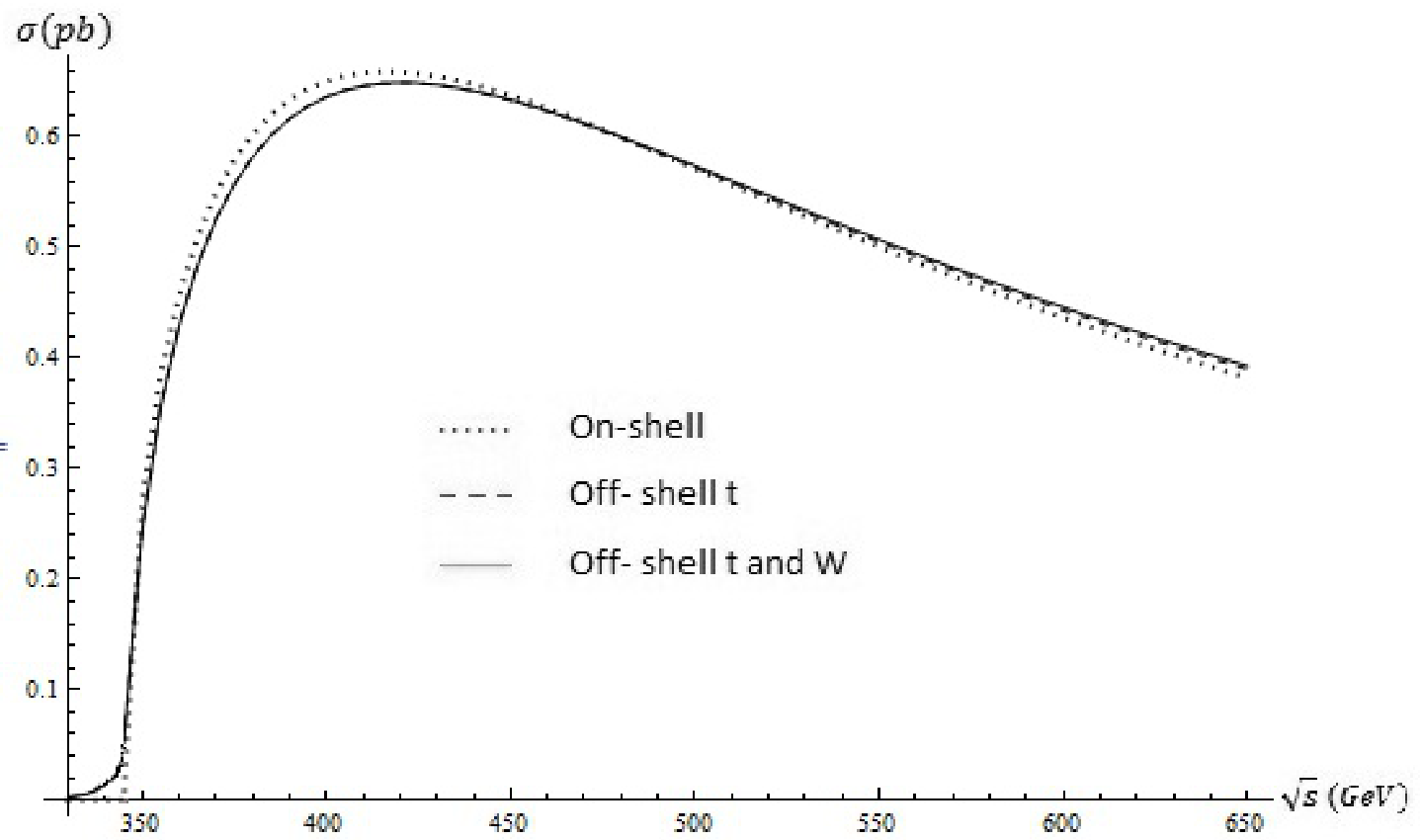}}
\caption{The cross-sections of the processes $e^+e^-\to t\bar{t}$,
$e^+e^-\to b\bar{b}W^+W^-$ and $e^+e^-\to b\bar{b}\sum_f4f$.}
\label{fig2}
\end{figure}

The results of the SMUP model calculations are presented in
Fig.~\ref{fig2}. Here, the dotted line represents the cross-section
of the top-pair production in the stable particle approximation
(SPA), i.e. without smearing of the top mass. The dashed line is the
cross-section incorporating the top mass smearing or top
finite-width effects (FWE) only, and the solid line gives the full
mass smearing result, both top-quarks and $W$-bosons. Note, that the
second case corresponds to the standard treatment of the process
$e^+e^-\to t^*\bar{t}^*\to b\bar{b}W^+W^-$ in the stable $W$-boson
approximation, and the third case -- to the full process shown in
Fig.~\ref{fig1}.

From Fig.~\ref{fig2}, one can see that the contribution of the top
quarks' FWE's is significant (up to a few percents in the
near-threshold region), while the contribution of $W$-bosons' FWEs
is small. The comparison of our results with ones in the standard
perturbative treatment shows that deviations are typically very
small. For instance, it was obtained in Ref.~[\refcite{17}], that
$\sigma(e^+e^-\to t^*\bar{t}^*\to b\bar{b}W^+W^-)$ for
$\sqrt{s}=500\,\mbox{GeV}$ is equal to 629 fb for
$M_t=150\,\mbox{GeV}$ and 553 fb for $M_t=180\,\mbox{GeV}$. For the
same input data, we have obtained 630 fb and 554 fb, respectively,
which are in a good agreement with the result mentioned above. This
comparison proves the applicability of the SMUP model fermion
propagator given by the first expression in Eq.~(\ref{2.4}). In
Section 4, we make such a comparison for exclusive processes as well
where both SMUP model fermion and boson propagators Eq.~(\ref{2.4})
are used.

It should be noticed also that we consider the FWE's, which are
significant in the near-threshold region, but we do not include
near-threshold effects caused by possible intermediate $t\bar{t}$
bound states. Since the top mean lifetime is considerably shorter
than the hadronisation time, the bound state effect has no sharp
resonant nature. However, it can be comparable with FWE's or
mass-smearing effects under consideration, and this problem will be
considered in more detail elsewhere.

\section{Factorizable corrections to the cross-section}

As it was shown in previous papers [\refcite{10}], the EW and QCD
corrections give large contributions to the cross-section of the
top-pair production at energy scales close to its threshold. In this
Section, we describe the strategy of our model calculations and give
the total cross-section including the principal part of NLO EW and
QCD corrections. Note, that the strategy of calculations and the
choice of input parameters are mainly caused and defined by the
effective character of the model treatment. In the framework of the
SMUP model, the instability (or finite width) of unstable particles
is accounted for by the smearing of their masses, i.e. by the
probability density function $\rho(m)$. In turn, this function
contains momentum dependent parameters $M(q)$ and $\Gamma(q)$ in
analogy with the standard perturbative treatment which uses dressed
propagators. So, in that sense the corrections of self-energy type
are already included at the ``effective'' tree level, and it is
reasonable to use an effective couplings, such as running coupling,
absorbing the major part of vertex-type corrections.

In our calculations we have used the following input data
[\refcite{18}]:
\begin{align}\label{3.1}
&\alpha(M_Z)=0.00781763,\,\,\,\alpha_s(M_Z)=0.118,\,\,\,\sin^2\theta_W(M_Z)=\hat{s}^2_Z=0.2313,\notag\\
&M_Z=91.1876\,\mbox{GeV},\,\,\,M_W=80.399\,\mbox{GeV},\,\,\,M_t=172.9\,\mbox{GeV}.
\end{align}
The running coupling constants $\alpha_k(Q^2),\,k=1,2,3$ were used
in the one-loop approximation:
\begin{equation}\label{3.2}
\alpha_k(Q^2)=\frac{\alpha_k(M_Z)}{1-(\beta_k/2\pi)
\ln(Q^2/M^2_Z)},\,\,\,\beta_k=(4.1,\,-19/6,\,-7).
\end{equation}

The cross-sections are calculated including the following
corrections:
\begin{itemize}
\item Vertex and self-energy type corrections for stable particles
are mainly included into running couplings (\ref{3.2}).\\
\item Self-energy corrections for unstable particles are included
into the probability density function $\rho(m)$, which describes
the smearing of UP's masses.\\
\item Initial state radiation (ISR) is described by the photon radiation spectrum
[\refcite{19,20}], and the bremsstrahlung from the final $t$-quark
states
-- by vertex $Q$-dependent factor [\refcite{21}].\\
\item QCD corrections to the top production and decay are described
by the vertex multiplicative factor [\refcite{21}].\\
\item Contribution of the box diagrams to the total cross section
was evaluated at energy scales close to the threshold by using
numerical FormCalc v7.3 [\refcite{22}] routines.
\end{itemize}

\begin{figure}
\centerline{\includegraphics[width=.6\textwidth]{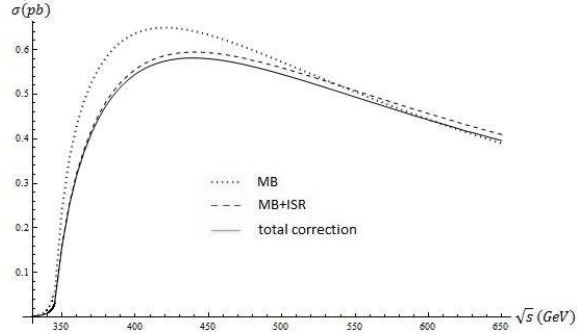}}
\caption{The higher order corrected cross-sections of the process
$e^+e^-\to b\bar{b}\sum_f4f$.} \label{fig3}
\end{figure}

The higher order corrected cross-sections of the inclusive process
$e^+e^-\to t^*\bar{t}^*\to b\bar{b}\sum_f4f$ are shown in
Fig.~\ref{fig3}. There, the dotted line represents the Born model
cross-section, the dashed line -- the cross-section with ISR and the
solid line -- the cross-section with total factorisable corrections
(without box diagrams contribution). From the figure, one can see
that the main contribution is given by ISR correction, which
significantly reduces the cross-section in the near-threshold energy
range and increases it at energy scales above $\sim$0.6 TeV. At
large energies ($\sqrt{s}>0.5\,\mbox{TeV}$) the contribution of EW
and QCD corrections becomes significant and has to be properly taken
into account.

\begin{figure}[h!]
\centerline{\epsfig{file=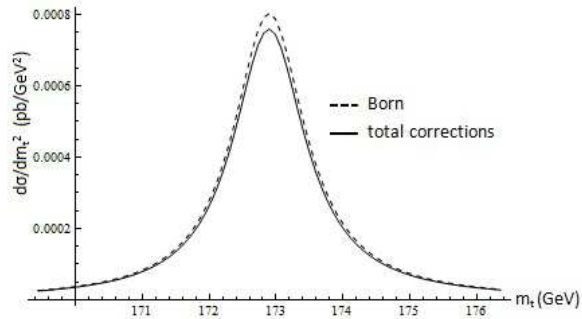,width=9cm}}
\caption{Invariant top mass distribution.} \label{fig4}
\end{figure}

\begin{figure}[h!]
\centerline{\epsfig{file=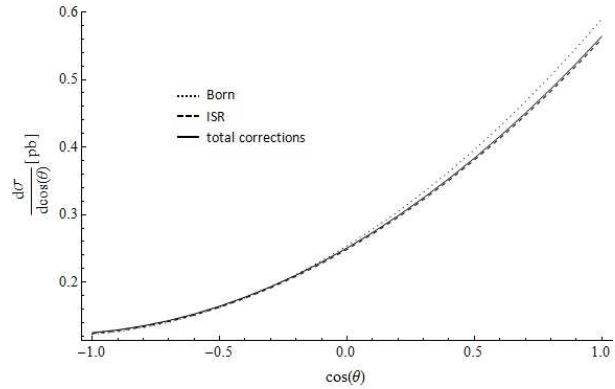,width=8cm}} \caption{The
angular differential cross-sections for the process $e^+e^-\to
b\bar{b}\sum_f4f$.} \label{fig5}
\end{figure}

In Fig.~\ref{fig4} we present the invariant mass distribution and
illustrate the influence of various corrections on it. One can see
that the corrections which we have taken into account according to
the procedure above give noticeable contribution into this
distribution in the peak area. We, also, illustrate such influence
on the angular differential cross-section presented in
Fig.~\ref{fig5}. Again, we notice that this influence is quite
significant and should be taken into account.

\section{The cross-sections of exclusive processes}

So far, we have considered the cross-section of inclusive process
$e^+e^-\to t^*\bar{t}^*\to b\bar{b}\sum_f4f$ where the final state
is summed up over all possible fermion flavors. As was noticed in
the second Section, in order to get the cross-section of exclusive
process $e^+e^-\to t^*\bar{t}^*\to b\bar{b}f_1f_2f_3f_4$ we can
include the corresponding branching ratios $\mathrm{Br}(W\to
f_1f_2)$ and $\mathrm{Br}(W\to f_3f_4)$. Acting this way we obtain
\begin{equation}\label{4.1}
\sigma(e^+e^-\to b\bar{b}f_1f_2f_3f_4)=\sigma(e^+e^-\to
b\bar{b}\sum_f4f)\mathrm{Br}(W\to f_1f_2)\mathrm{Br}(W\to f_3f_4)\,.
\end{equation}
where we omit intermediate virtual $t^*\bar{t}^*$ state for
simplicity. This relation directly follows from the Eq.~(\ref{2.3})
when one substitutes a partial decay width of the $W$-boson into
numerator of the probability distribution function $\rho_W(m)$
instead of the total width. It can be also derived by
straightforward calculation of the $\sigma(e^+e^-\to
b\bar{b}f_1f_2f_3f_4)$ in the framework of the effective theory (see
Ref.~[\refcite{12}]).

The expressions for the branchings ratios $\mathrm{Br}(W\to f_1f_2)$
were considered in detail in Ref.~[\refcite{21}]. Here, we use very
simple but sufficiently precise formulae which incorporate QCD
corrections:
\begin{equation}\label{4.2}
\mathrm{Br}(W\to
l\bar{\nu}_l)=\frac{1}{9(1+2\alpha_s(M_Z)/3\pi)},\,\,\,\mathrm{Br}(W\to
u_i\bar{d}_k)=\frac{|V_{ik}|^2(1+\alpha_s(M_Z)/\pi)}{3(1+2\alpha_s(M_Z)/3\pi)},
\end{equation}
where $V_{ik}$ are elements of the Cabibbo-Kobayashi-Maskawa mixing
matrix. We, also, employ the QCD corrected expression for the top
quark width [\refcite{21,23}]:
\begin{equation}\label{4.3}
\Gamma(t\to
bW)=\frac{1}{16}\alpha_2(M_t)|V_{tb}|^2\,\eta_{QCD}\,M_t\,
f(M_t,M_W,M_b),
\end{equation}
where
\begin{eqnarray}\label{4.4}
&&f(M_t,M_W,M_b)=\lambda(M^2_b,M^2_W;M^2_t)\left(\frac{(M^2_t-M^2_b)^2}{M^2_t
M^2_W}
+\frac{M^2_t+M^2_b-2M^2_W}{M^2_t}\right);\\
&&\lambda(M^2_b,M^2_W;M^2_t)=\left(1-2\frac{M^2_b+M^2_W}{M^2_t}
+\frac{(M^2_W-M^2_b)^2}{M^4_t}\right)^{1/2};\nonumber \\
&&\eta_{QCD}=1-\frac{2\alpha_s(M_t)}{3\pi}\left(\frac{2\pi^2}{3}-\frac{5}{2}\right).
\nonumber
\end{eqnarray}

Using Eqs.~(\ref{4.1})--(\ref{4.3}) we can calculate the exclusive
cross-section for an arbitrary six-fermion final state
$(b\bar{b}f_1f_2f_3f_4)$. Such calculations taking into account the
factorizable EW corrections were performed within the standard
perturbative approach for the case of
$(b\bar{b}\mu^+\nu_{\mu}\mu^-\bar{\nu}_{\mu})$ and
$(b\bar{b}\mu^+\nu_{\mu}d\bar{u})$ final states in
Ref.~[\refcite{10}]. In this work, the full set of topologically
different Born diagrams leading to the same six-fermion final state
was considered. It was shown, that certain cuts on invariant masses
of the $Wb$ and $f_if_k$ pairs, which correspond to intermediate
$t,\bar{t}$ and $W^+,W^-$ states for signal diagrams, significantly
reduce the relative contribution of the background (see
Table~\ref{tab1}).

In Tables~\ref{tab1} and \ref{tab2} the cross-sections are given for
two distinct reactions
\begin{equation}\label{4.5}
 (1):\quad e^+e^-\to b\bar{b}\mu^+\nu_{\mu}\mu^-\bar{\nu}_{\mu},\qquad (2):\quad e^+e^-\to
b\bar{b}\mu^+\nu_{\mu}d\bar{u}.
\end{equation}
for the energies $\sqrt{s}=430,\,500,\,1000,\,\mbox{GeV}$. In
Table~\ref{tab1} the cross-sections are presented in the Born
approximation for total set of diagrams
($\sigma_{Born}^{(k)}$(total)) and for the signal diagrams
($\sigma_{Born}^{(k)}(t^*\bar{t}^*)$), where $k=1,2$ denotes the
first and second reactions in Eq.~(\ref{4.5}), respectively. These
values (in fb) are taken from Table 1 in Ref.~[\refcite{10}] and are
calculated with the kinematical cuts $\delta_i<0.1$, where
$\delta_i$ is the deviation of the ratio $m_i^{inv}/M_i$ from unity
and index $i$ is related to different $t,\bar{t},W^+,W^-$ states
(for more details, see Ref.~[\refcite{10}]).
\begin{table}[!h] \small
\tbl{Born-level cross-sections of the processes $(1)$ and $(2)$ in
Eq.~(\ref{4.5}).}
{ \begin{tabular} {@{}|c|c|c|c|c|@{}}    
\hline
 $\sqrt{s}$, GeV
 &$\sigma_{Born}^{(1)}$(total)&$\sigma_{Born}^{(1)}(t^*\bar{t}^*)$&
 $\sigma_{Born}^{(2)}$(total)&$\sigma_{Born}^{(2)}(t^*\bar{t}^*)$\\
\hline
  430&5.9117&5.8642&17.727&17.592\\ \hline
  500&5.3094&5.2849&15.950&15.855\\ \hline
  1000&1.6387&1.6369&4.9134&4.9106\\
\hline
 \end{tabular} \label{tab1} }
\end{table}

In Table~\ref{tab2} the results for the total cross sections (in fb)
of processes $(1)$ and $(2)$ from Eq.~(\ref{4.5}) in the Born
approximation are shown in the second column. The cross-sections
with separate ISR and factorizable EW (FEWC) corrections are
presented in the third and forth columns, respectively, and the
cross-section with both the FEWC and ISR corrections included -- in
the fifth column. All values are calculated with the kinematical
cuts mentioned above.
\begin{table}[!h] \small
 \tbl{Comparison of the exclusive cross-sections of Ref.~[11] 
  and ones obtained in the present work.}
{ \begin{tabular} {@{}|c||c|c|c|c|@{}} \hline
 $\quad\sqrt{s},\,\rm{GeV}\quad$&  $\quad\sigma^{t^*\bar{t}^*}_{\rm Born}\quad$ &
 $\quad\sigma_{\rm Born+ISR}\quad$ & $\quad\sigma_{\rm Born+FEWC}\quad$
 & $\quad\sigma_{\rm Born+ISR+FEWC}\quad$  \\
\hline
 \multicolumn{5}{|c|}{$e^+e^-\to b\nu_{\mu}\mu^+\bar{b}\mu^-\bar{\nu}_{\mu}$, Ref.~[\refcite{10}] }      \\ \hline
 430              &   $5.8642(45)$ & $5.2919(91)$ & $5.6884(55)$ & $5.0978(53)$   \\
 500              &   $5.2849(43)$ & $5.0997(51)$ & $4.9909(49)$ & $4.8085(48)$   \\
 1000             &   $1.6369(15)$ & $1.8320(18)$ & $1.4243(14)$ & $1.6110(16)$   \\
\hline
 \multicolumn{5}{|c|}{$e^+e^-\to b\nu_{\mu}\mu^+\bar{b}\mu^-\bar{\nu}_{\mu}$, this work}        \\ \hline
 430              &   $5.86476$  & $5.27613$ & $5.77727$ & $5.19941$     \\
 500              &   $5.27352$  & $5.08651$ & $5.18407$ & $5.00291$     \\
 1000             &   $1.63061$  & $1.83508$ & $1.58925$ & $1.79079$     \\
\hline\hline
 \multicolumn{5}{|c|}{$e^+e^-\to b\nu_{\mu}\mu^+\bar{b}d\bar{u}$, Ref.~[\refcite{10}] }      \\  \hline
 430              &   $17.592(13)$  & $15.857(20)$ & $17.052(16)$ & $15.283(16)$   \\
 500              &   $15.855(13)$  & $15.311(15)$ & $14.977(16)$ & $14.438(14)$   \\
 1000             &   $4.9106(46)$  & $5.4949(55)$ & $4.2697(40)$ & $4.8287(47)$   \\
\hline
 \multicolumn{5}{|c|}{$e^+e^-\to b\nu_{\mu}\mu^+\bar{b}d\bar{u}$, this work}                     \\  \hline
 430              &   $17.8163$ & $16.0351$ & $17.5540$ & $15.8019$    \\
 500              &   $16.0203$ & $15.4517$ & $15.7516$ & $15.1979$    \\
 1000             &   $4.95397$ & $5.57465$ & $4.82889$ & $5.44011$    \\
\hline
\end{tabular} \label{tab2} }
\end{table}

From Table~\ref{tab2}, it follows that the differences of the model
and standard Born cross-sections are of an order of 0.1 percent and
ISR correction increases it only slightly. In principle, these
deviations can be further reduced. The situation becomes worse, when
we take into account all major corrections. The deviations increase
and become up to a few percents. This discrepancy is caused by the
fact that in Ref.~[\refcite{10}] an additional contribution from the
non-signal (background) diagrams was included while we consider the
signal contribution only. Moreover, we do not include the
contribution of the box diagrams which becomes very important at
large energies far from the threshold. According to estimations in
the framework of the standard perturbative treatment, the box
diagrams contribution is of an order of a few percents in the
near-threshold energy range. Rough estimations in the framework of
the SMUP model give the box contribution equal to $1.5 - 2$ percents
in the energy region under consideration, and these estimations
decrease the deviations. However, in the framework of the SMUP
model, as well as in the effective theory of UP, the higher order
corrections have an effective character, and a consistent
formulation of the perturbative treatment with this model is
required. This problem leads to using the model propagators inside
loop diagrams, but the validity of such a procedure has still to be
justified theoretically. In particular, one should first analyze the
asymptotic properties of the propagators. The analysis is not
carried out yet, but it is in progress now. Note, that a good
agreement of the SMUP model and standard Born-level results,
illustrated in Table~\ref{tab2}, provides a good basis for such an
analysis.

\section{Conclusion}

The production of the $t\bar{t}$ pair and its subsequent decay into
six fermion final states in $e^+e^-$ annihilation has been
previously analyzed within the standard perturbative treatment in a
vast literature. In this work, we performed the corresponding
analysis in the framework of SMUP model. So far, this approach was
applied mainly to the gauge boson production, where the structure of
the model boson propagators was successfully tested
[\refcite{13,14,14a}]. In the present work, we have tested the
structure of the model fermion propagator, and the top quark
production mechanism has been chosen as an important example. It was
shown that the results of Born-level calculations are in a good
agreement with the standard perturbative ones, providing the
applicability of the SMUP approach to the top-quark production and
decay processes.

The SMUP model provides simple analytical expressions for the total
cross-sections of inclusive and exclusive processes with top quark
pair production and its subsequent decay. It is a convenient and
simple instrument for description of complicated multi-step
processes with unstable particles participation. The precision of
this approach at the tree level is of an order of 0.1 percent or
better. The method gives a possibility to include, in principle, all
factorizable corrections. Our approach can be useful in a
preliminary analysis of complicated processes with intermediate
time-like top quark exchanges within the Standard Model and beyond.

\end{document}